# Analysis of AeroMACS Data Link for Unmanned Aircraft Vehicles

Maede Zolanvari, *Student Member, IEEE,* Marcio A. Teixeira, *IFSP, FAPESP*, Raj Jain, *Life Fellow, IEEE*

***Abstract*****—Aeronautical Mobile Airport Communications System (AeroMACS) is based on the IEEE 802.16e mobile wireless standard commonly known as WiMAX. It is expected to be a main part of next-generation aviation communication system to support fixed and mobile services for manned and unmanned applications. AeroMACS will be an essential technology helping pave the way toward full integration of Unmanned Aircraft Vehicle (UAV) into the national airspace. A number of practical tests and analyses have been done so far for AeroMACS. The main contribution of this paper is to consider the theoretical concepts behind its features and discuss their suitability for UAVs' applications. Mathematical analyses of AeroMACS physical layer framework is provided to show the theoretical trade-offs. We mainly focus on the analysis of AeroMACS OFDMA structure, which affects the speed limits, coverage cell, channel estimation requirements and inter-carrier interference.**

## I. INTRODUCTION

Aeronautical Mobile Airport Communications System (AeroMACS) is based on IEEE 802.16e standard, commonly known as, WiMAX. AeroMACS was developed as a solution for the congested VHF spectrum at the airports [1]. It was developed primarily for stationary wireless communication; later it was adapted for mobile wireless communication as well. Global interoperability is the main object of all future aeronautical networks. AeroMACS, as an important element of Future Communication Infrastructure (FCI), is supposed to be a part of providing seamless communication worldwide. AeroMACS, developed initially at Radio Technical Commission for Aeronautics (RTCA), has been internationally standardized as an air interface through different related organizations such as International Civil Aviation Organization (ICAO) and European Organization for Civil Aviation Equipment (EUROCAE).

AeroMACS has been introduced as one of the most powerful candidates to be used as data link communication for on-airport communication, including both manned and unmanned aircraft. Providing a proper positioning system for Unmanned Aircraft Vehicles (UAVs) is essential, due to the absence of the pilot and being deprived of a direct vision [2]. AeroMACS provides an accurate low latency positioning system that can be used by UAVs. Other benefits of using AeroMACS is for navigation, security, air traffic, weather, and emergency information.

As the aircraft (manned or unmanned) may pass through different countries' borders or continents' airspace, service agreements, access, and authorization in air traffic control must be managed on a global scale. The global standardization must consider all the various types of aviation data links such as satellite systems, cellular systems, and future wireless systems. Two main programs were created for this purpose, Next Generation Air Transportation System (NextGen) in the United States and Single European Sky ATM (Air Traffic Management) Research (SESAR) [3]. AeroMACS is a big part of these programs as a proper candidate for aviation. A European-based project called Seamless Aeronautical Networking through the integration of Data Links Radios and Antennas (SANDRA) and the IEEE WiMAX Standard Committee are working on different configuration options to improve the performance of AeroMACS.

AeroMACS uses the frequency range of 5.091-5.150 GHz, C-Band, Aeronautical Mobile (Route) Service (AM(R)S) allocation. Channel bandwidth is 5 MHz, operating in time division duplex (TDD) mode along with the orthogonal frequency division multiple access (OFDMA) frame structure. It can provide throughput up to 10 Mbps over a range of 3 km.

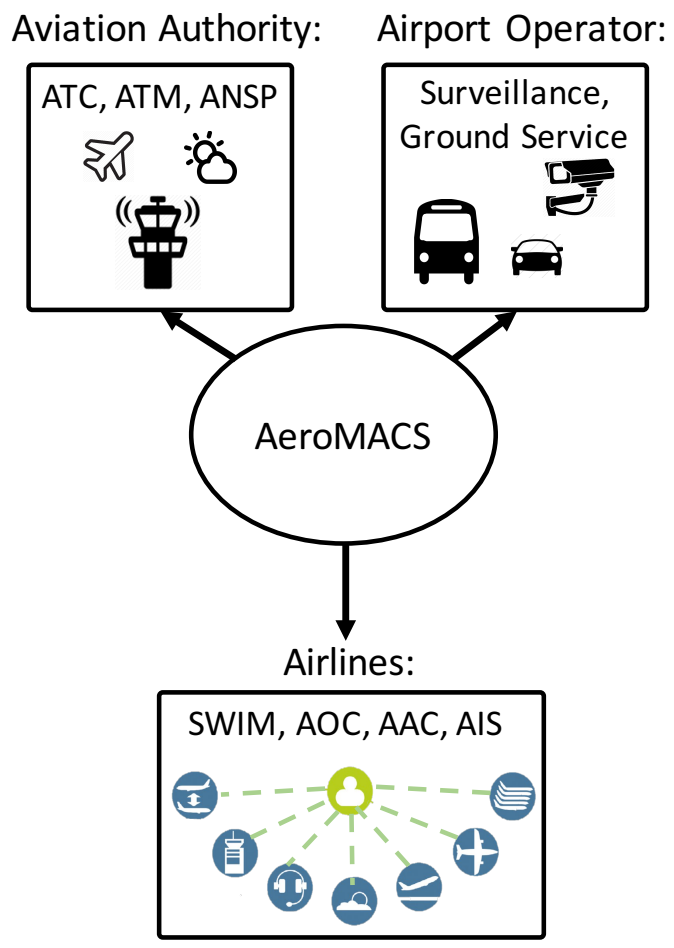

Figure 1. AeroMACS Applications

Some of the AeroMACS applications are shown in Fig. 1. It can be used by aviation authorities for applications such as Air Navigation Service Providers (ANSP), Air Traffic Control (ATC) and Air Traffic Management (ATM). In the airports, it can be employed for live surveillance using fixed or portable cameras. Further, AeroMACS can be used as a communication system for ground services such as emergency, security, fire, and rescue operations. AeroMACS is the only recognized technology that can support all these services simultaneously and through only one infrastructure [4]. AeroMACS is planned to be a part of System-Wide Information Management (SWIM). SWIM is an advanced technology programmed by the Federal Aviation Administration (FAA) to provide a secure platform for cooperation among the national and international aviation organizations.

### A. AeroMACS System

Fig. 2 shows the main components of an AeroMACS system. MS is the Mobile Station (e.g., airplanes, surface vehicles), and GS is the Ground Station (e.g., ATC base station). MSs are WiMAX Customer Premise Equipment (CPE), and GSs are WiMAX base stations, both designed using the AeroMACS profile [5]. The GS-mngt is the ground station management. It is usually a software set up on a personal computer to manage and configure the GS. AAA is the Authentication, Authorization and Accounting server. Authentication ensures that the users are who they claim to be. Authorization ensures that they can access only permitted resources. Access Service Network Gateway (ASN GW) is the router to the external IP network.

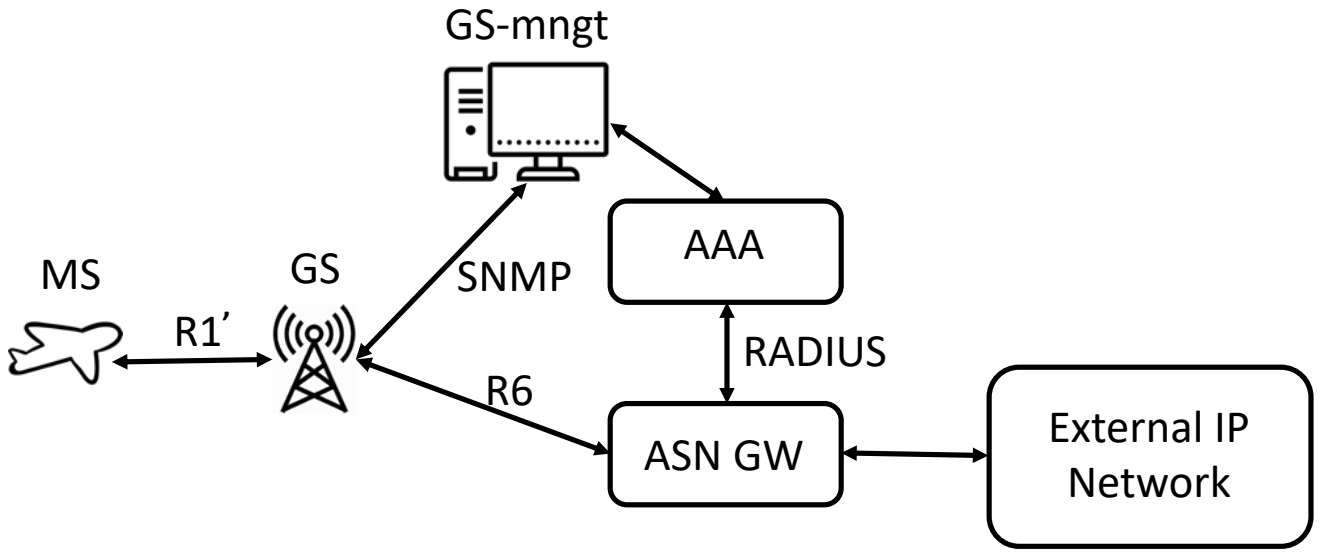

Figure 2. AeroMACS system overview

The main interfaces are R1' and R6. R1' is the R1 interface of IEEE 802.16e with AeroMACS profile adaptation for communications between the MS and the GS. R6 is the standard IEEE 802.16e interface between the GS and the ASN GW. Remote Authentication Dial-in User Service (RADIUS) is the commonly used protocol for AAA communications designed by the Internet Engineering Task Force (IETF). And finally, Simple Network Management Protocol (SNMP) is an Internet-based protocol for monitoring and managing all devices in the network. AeroMACS is the wireless data link between the aircraft and the Air-Ground Communication Service Provisioning (ACSP) system. The ACSP consists of the ground station, ASN GW, AAA server and routing infrastructure.

Performance requirements for the data links of air traffic data communication are set by EUROCAE Working Group (WG) 78 or WG-78 as the bounds on Transaction Time (TT). The TT at 99.9% (TT99.9) or the TT at 95% (TT95) is the time when, 99.9% or 95% of all transactions, respectively, are completed. TT is related to the continuity constraints, and continuity is related to the efficiency of the data link. The minimum required TT in ACSP should be less than 9 s for 99.9% and less than 4 s for 95% of the messages. ACSP of AeroMACS is designed with the goal of 500 ms for TT95, and 1 s for TT99.9, which are lower than minimum requirements. Being able to prioritize the services of AeroMACS usually provides even better transaction time performance for important tasks. This feature makes AeroMACS a unique data link among other communication systems being used or planned for future use in aviation so far [6]. This feature can help UAVs in mission-critical tasks, since the communication delay of the data link will not be a problem anymore.

The rest of this paper is organized as follows. We provide the state of the art of prior analysis of AeroMACS in Section II. After that, we discuss the different features of AeroMACS and how they might be improved in Section III. Section IV provides an analysis of the OFDMA structure, followed by a summary in Section V.

## II. PRIOR WORKS

Bartoli et al. [7] provide a comprehensive survey of AeroMACS related to its physical and MAC layer properties. Features such as synchronization issues, handover techniques, and resource management are discussed. Some general ideas on network architecture and minimum requirements for AeroMACS are also studied.

Pulini et al. [8] focus on the synchronization and channel estimation techniques used in AeroMACS. The overall error rate performance during the approach, landing, and takeoff phases of an aircraft flight are analyzed. To improve the performance regarding robustness against synchronization errors, several modifications on conventional methods, already used in these systems, are suggested.

Kerczewski et al. [9] provide the results of practical tests conducted from 2009 through 2012 on an AeroMACS prototype at the NASA-CLE Communications, Navigation and Surveillance (CNS) Testbed. They briefly talk about the status of the AeroMACS standardization process. Different standardization-related organizations that are active in the AeroMACS improvement such as RTCA, WiMAX Forum, and ICAO along with their goals and plans are discussed.

In a second paper, Kerczewski et al. [10] discuss applying IEEE 802.16j multi-hop relays standard to the AeroMACS prototype. With relays, the direct link between the GS and MS is replaced by a GS to Relay Station (RS) link followed by an RS to MS link. This multi-hop protocol scheme is shown to increase the AeroMACS system's capacity and reduce the interference and the path loss.

Kamali et al. [11] improve the relay scheme analyzed by Kerczewski et al. in [10] even further. The two relay modes supported by IEEE 802.16j modification are applied to the proposed AeroMACS multi-relay scheme. The relays can choose to operate in transparent mode or non-transparent mode. Advantages and disadvantages of both are mentioned through the impacts of relay type selection. The effects of applying centralized or distributed communications resource management and scheduling procedures are investigated. In the centralized technique, every decision is made by the GS, whereas in the distributed method, relay nodes are responsible for the resource sharing. Choosing either of the two schemes affects network parameters such as latency and signal overhead of PHY and MAC layer protocols. Changing these parameters would lead to different throughput and Quality of Service (QoS).

Correas and Fistas [12] show that some AeroMACS services can benefit from a flexible asymmetric ratio of the number of OFDMA symbols assigned to downlink (DL) and uplink (UL) channels. Their research is based on the fact that AeroMACS's TDD framework support different shares of throughput between DL and UL. They provide a comprehensive analysis of the comparison of different DL/UL symbol ratios that can be used in an airport environment. The examined ratios are based on the cell constraints and data rate requirements. The main focus of the paper is on applications such as video surveillance and sensors using the AeroMACS system.

Morioka et al. [13] study the placements of onboard UAV antennas using AeroMACS as the command and non-payload communications (CNPC) link. The main objective of this paper is to find optimal antenna placement since its positional relationship with the ground station's communication antenna is not fixed. Due to their flight tests, placing the multiple input multiple output (MIMO) antennas on the UAV in vertical position showed the best performance.

Plass et al. [14] investigate the compatibility of the European-funded project Seamless Aeronautical Networking through Integration of Data Links, Radios, and Antennas (SANDRA). The principal objective of the paper is to study integrating currently used data links, VDL2, and satellite-based BGAN links, with AeroMACS communication link. Through seamless layer 2 and layer 3 handovers, it is shown that the SANDRA network is flexible, compatible and scalable.

Currently, the United States and Europe are working together to standardize AeroMACS for aviation integrating both unmanned and manned aircraft. In the United States, RTCA Special Committee (SC) 228 is setting Minimum Operational Performance Standards (MOPS) for data links for unmanned aircraft. In Europe, EUROCAE WG 82 is also developing MOPS. The ICAO is working on Standards and Recommended Practices (SARPS) and Concept of Operations (ConOps) for unmanned aerial systems. ICAO has planned to come up with a common standard for airport communication in the United States and Europe while using AeroMACS as a part of this plan for the global communication purposes [9]. In the next sections, we focus on several aspects of AeroMACS that must be analyzed to understand its limitations.

## III. AEROMACS FEATURES ANALYSIS

In this section, we discuss the different features of AeroMACS and their suitability for UAV's applications. AeroMACS has several characteristics that make it a strong potential candidate for future global aviation data links. It is expected to increase the safety, adaptability, and efficiency of airport communications. However, like all other standards, there are some limitations associated with it. In the following sub-sections, we discuss AeroMACS features such as adaptiveness, physical layer, and data link layer.

### *A. Adaptiveness*

AeroMACS provides link adaptation and dynamic bandwidth allocation. These are especially useful in environments that suffer from fast fading. By using the fast feedback channel allocation (FFCA), MS can report the channel quality and Signal to Noise ratio (SNR) to the GS. The GS changes the resource allocation depending upon the link quality. This feature is very promising due to the different environmental situations that the UAV's data link might experience based on the application.

The layer 2 of the AeroMACS supports different QoS through different performance characteristics, such as delay, jitter, packet loss, and throughput. On the physical layer (PHY), AeroMACS system can select among different Modulation and Coding Schemes (MCS): QPSK, 16-QAM, 64-QAM, with different forward error correction (FEC) ratios such as 1⁄2, 2⁄3, 3⁄4 and 5⁄6. Theoretically, the operating range of QPSK is two times larger than 16-QAM, and the operating range of 16-QAM is two times larger than the operating range of 64-QAM. Theoretically, halving the number of bits per symbol can double the transmission range, but the expected throughput becomes half. However, in practice, due to bit errors, using larger modulation constellations does not exactly double the throughput. Based on the results in [15], Table I shows the measured results of AeroMACS throughput for three different modulation schemes with the same coding rate, 1⁄2. The throughput is highly dependent not only on the modulation scheme but also on the coding rate, radio frequency (RF) channel conditions and even on the distance between the MS and the GS.

TABLE I
AEROMACS EXPECTED THROUGHPUTS VS MODULATION SCHEMES

| **Modulation Coding Scheme** | **Downlink (kbps)** | **Uplink (kbps)** |
|---|---|---|
| ***QPSK 1/2*** | 983.3 | 532.4 |
| ***16-QAM 1/2*** | 2153.52 | 1235.52 |
| ***64-QAM 1/2*** | 3595.04 | 1758.48 |

Being able to choose among different modulation types and FEC ratios can ensure a very strong data link for the UAV that will not be affected by the surrounding's conditions. For instance, UAVs used for delivery or transportation applications would pass through different zones, some of which might be condensed urban areas with high levels of radio interference. Changing the PHY attributes dynamically based on the UAV's path is a great bonus of AeroMACS-based data links for UAV's communications.

*B. Physical Layer*

AeroMACS uses 2×2 MIMO with 5 ms OFDMA frames. The choice of 2×2 MIMO is optimal. The number of antennas is not so large that it would make the equipment heavy or impractical to use it on small UAVs. Also, the computation complexity for combining the received signals from only two antennas is not very high. Further, 2×2 MIMO provides sufficient physical diversity to help with fading or interference problems, which are significant challenges in UAV aviation. The antennas are dual-slant antennas using polarities offsets of +45° and -45° degrees from the Horizontal and Vertical axes. Hence, the receiver would see 100% of an H/V transmitted signal.

Regarding the GS specific physical layer characteristics, the directional antenna gain is about 15 dBi. The typical GS operating temperature is from -40°C to +55°C, which is sufficient to operate in a wide range of different weather conditions. The typical power consumption of the GS is about 75 W, with maximum antenna transmit power equal to 2 x 23 dBm. The multiplication with 2 comes from the fact that a 2×2 MIMO antenna is used.

MS uses an omnidirectional antenna with 6 dBi gain, and the power consumption is about 12 W, which is relatively a bit high compared to the users in traditional communication systems. The MS dimensions should be at least 300×300×90 mm to carry the equipment. Typical operating temperature is -10° to 55°C, where -10°C may not be functionally sufficient considering the height of the aircraft; especially in cold weathers, it is not enough at all.

Some applications of UAVs must perform constantly during the year. Using UAVs for monitoring and surveillance is an example of these applications. These UAVs would experience very cold weather during winter especially in northern latitudes. Further, UAVs helping in search and rescue missions during a disaster might have to operate in harsh weather situations. If AeroMACS is chosen to be implemented as UAV's data link for these applications, proper solutions must be taken into considerations, since -10°C operating temperature is not sufficient enough.

AeroMACS uses GPS (Global Positioning System) for time synchronization. To mitigate the interference between AeroMACS cells, AeroMACS uses TDD structure, as mentioned before. Hence, all the GSs installed in the same zone will be synchronized with GPS time or any other time source having equivalent performance as GPS. All GSs within the same area should follow the same frame structure [16], which means that their uplink and downlink subframes are the same size and start at the same time from all nearby GSs.

*C. Data Link's Features*

Data link latency requirements of AeroMACS depend on several criteria. The time division between the downlink (i.e., from GS to MS) and the uplink (i.e., from MS to GS) messages is applied by the Human Interaction Time (HIT) based on the Required Communication Technical Performance (RCTP) and latency constraints. The RCTP is defined as the required 95$^{th}$ percentile latency for aeronautical data link technologies, and it is specified in the Communications Operating Concepts and Requirements (COCRv2) document. COCRv2 was a joint program by EUROCONTROL and FAA started in 2007 to define constraints on one-way latency (which is called TT95-1 way), continuity, integrity, and availability [17].

Flexible DL/UL time divisions offer flexible DL/UL data rates and throughput. A wide range of UAVs' applications can benefit from this feature of AeroMACS. Some applications require higher data rates for DL, and some require higher data rates for UL. For instance, in surveillance applications, UAV needs high data rates and throughput for UL to be able to send real-time videos. However, for applications such as delivery, the UAVs must be controlled and guided through the GS to finish the mission, which implies higher throughput required on DL.

Regarding the suitable frequency band for the data link used at the airports, Budinger and Hall [18] discuss previous NASA research on AeroMACS measurements at JFK Airport. A large number of power delay profiles (PDPs) and received signal strength (RSS) were measured to model different features of the radio channel of an airport environment. That study showed for more than 1 MHz bandwidth allocated to the data links used on that airport, the channel frequency starts getting dispersive. But during further studies, data requirements, and system tolerance, 5 MHz was chosen as the optimum bandwidth.

It is undeniable that wireless data links on airport surfaces are exposed to high levels of multipath fading and non-line-of-sight (NLOS) conditions. Despite all these challenges associated with the unique airports wireless channel's conditions, AeroMACS shows an adequate performance as a data link for aviation communications.

In most research that has been done so far on AeroMACS, the results come from experiments and running tests on an actual airport surface or testbed. In the following section, we show the trade-offs that different parameters of OFDMA structure have on the associated AeroMACS system features.

## IV. OFDMA FRAME STRUCTURE ANALYSIS

OFDMA features of a transmitted symbol through AeroMACS standard is shown in Table II. As mentioned before, AeroMACS uses radio spectrum from 5095 MHz to 5150 MHz, which consists of 11 channels, each with 5 MHz bandwidth. Each channel in the uplink has 17 sub-channels, and 15 sub-channels in the downlink. The sub-channels in uplink are composed of 272, 136, and 104 subcarriers for data, pilot, and null respectively. The sub-channels in downlink are consist of 360, 60, and 92 subcarriers for data, pilot, and null respectively. Hence, each 5 MHz channel bandwidth in the uplink and the downlink has 512 subcarriers.

TABLE II
AEROMACS PHYSICAL LAYER PARAMETERS

| **Parameters** | **OFDMA** |
|---|---|
| ***Bandwidth*** | 5 MHz |
| ***FFT size*** | 512 |
| ***Cyclic Prefix*** | 1/8 Ts = 12.8 µs |
| ***Frame Size*** | 24 OFDM symbols |
| ***Symbol Time (Ts)*** | 102.4 µs |
| ***Subcarrier Spacing*** | ≈10 kHz |

In the following sub-sections, we analyze different parameters, such as subcarrier spacing, maximum coverage, inter-carrier interference, coherence time and cyclic prefix length.

### *A. Subcarrier Spacing*

Allocating 5 MHZ bandwidth to each AeroMACS channel and having 512 of OFDMA subcarriers (i.e., equal to the FFT size), the subcarrier spacing would be 10 kHz, regarding the equation (1).

$$\Delta f = BW/(N_s + 1) \tag{1}$$

In this equation, $BW$ is the bandwidth, and $N_s$ is the number of subcarriers. There is a trade-off between the number of subcarriers and subcarrier spacing. Having a larger number of subcarriers means ending up with smaller subcarrier spacing. As the subcarrier spacing decreases, the symbol duration increases, and we would end up with a symbol duration much larger than the delay spread. Therefore, theoretically increasing the number of subcarriers should benefit the performance in a sense that the system will be able to tolerate larger delay spreads. However, from an implementation point of view, as we increase the number of subcarriers and assigning subcarrier frequencies very close to each other, several problems arise such as Doppler shift. The Doppler shift is a crucial issue especially in UAV data links that causes Inter-Carrier Interference (ICI), which we will discuss in the next sub-section.

The receiver needs to be synchronized to the carrier frequency almost precisely; if not, even a small carrier frequency offset will cause a large frequency mismatch between the neighbor subcarriers. Interference and receiving noisy signals would degrade the performance significantly. Hence, as the number of subcarriers increases, higher levels of synchronizations is needed. If a precise synchronization is required at the receiver components, the cost of RF hardware would be very high. As a result, a reasonable trade-off between the subcarrier spacing and the number of subcarriers must be taken into account.

### *B. Maximum Coverage*

The maximum path loss allowed by link budget in AeroMACS receiver is about 128 dB [19, 8]. The relationship between maximum Line-of-Sight (LoS) coverage $d_{max}$, frequency $f$, speed of light $c$, and maximum allowed path loss $PL_{max}$ is as follows:

$$d_{max} = \frac{c}{4\pi f}\sqrt{PL_{max}} \tag{2}$$

Accordingly, based on AeroMACS specifications, we would have the maximum coverage around 12 km LoS, which is also the maximum theoretical coverage stated in IEEE 802.16 standards. However, achieving this range of LoS is almost impossible especially at airports. Further, as the number of users demanding wireless services is going incredibly higher every day, and node density is getting condenser, practical LOS length is decreasing.

All the airport areas are considered as almost non-LOS. As the practical experiments on AeroMACS have shown, for each 2 km increase in range, we will have an excess loss of 10-20 dB based on different propagation models [20]. That is why the cell range in AeroMACS is not designed to be any more than 3 km. More specifically, the cell size for gate areas is about 1.1 km, and for runway and taxiway areas is about 2.5 km. Each aircraft should have access to at least 2 and many times 3 GSs to increase availability.

This range of coverage will be enough for some UAV's applications such as monitoring, surveillance, and constructions. However, if AeroMACS is used as data link for UAVs in delivery or transportation, this range is not enough at all. In this case, a proper solution might be employing several AeroMACS GS on the way of the UAV to extend the coverage. However, challenges such as hand over issues must be resolved.

### *C. Inter-Carrier Interference*

The Doppler shift is caused due to the motion of the MS, which makes the received frequency at the GS and MS differ from the sent frequency. The difference may be positive or negative depending on whether the MS is getting closer to or further from the GS. Doppler shift, $f_D$, can be calculated using the equation (3) below:

$$f_D = \frac{v \times f}{c} \tag{3}$$

Here, $v$ is the velocity of the MS, $f$ is the frequency of the carrier, and $c$ is the speed of light.

We can calculate the power of ICI, $\sigma_I^2$, caused by Doppler shift using the equation (4) below:

$$\sigma_I^2 = E[|I(k)|^2] \\ = E_s - \frac{E_s}{{N_s}^2}\sum_{k=0}^{N_s-1}\sum_{k'=0}^{N_s-1} J_0(2\pi f_D T_s(k-k')) \tag{4}$$

Here, $E_s$ is the symbol energy, $J_0()$ is the zeroth-order

Bessel function of the first kind, $T_s$ is the symbol period, and $N_s$ is the number of subcarriers. We set $E_s$ equal to 24 dBm, which is the average maximum signal power per subcarrier in a typical AeroMACS communication system [21].

$\sigma_I^2$ is also known as the variance of ICI [22]. The graph of Doppler shift in kHz and the ICI in dBm for different speeds of MS have been shown in Fig. 3 and Fig. 4, respectively. Signal to ICI ratio vs. MS speed in km/h is also shown in Fig. 5. Doppler shift mostly affects the speed of the UAVs, which we will talk about later.

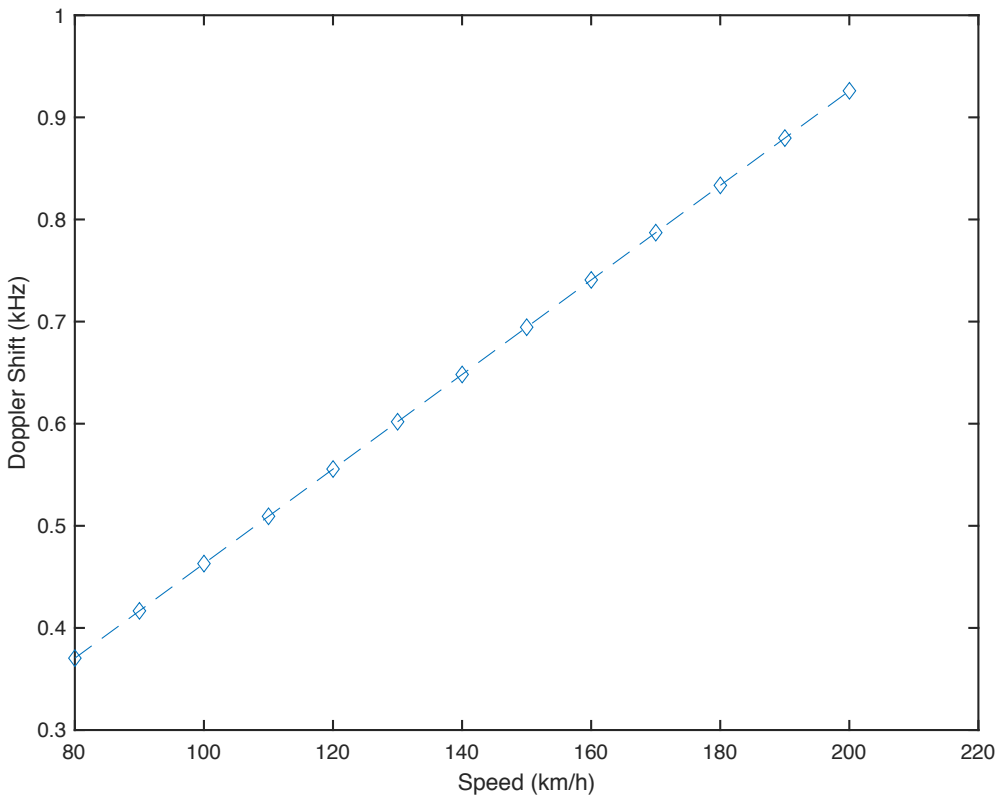

Figure 3. Doppler Shift caused by aircraft movement

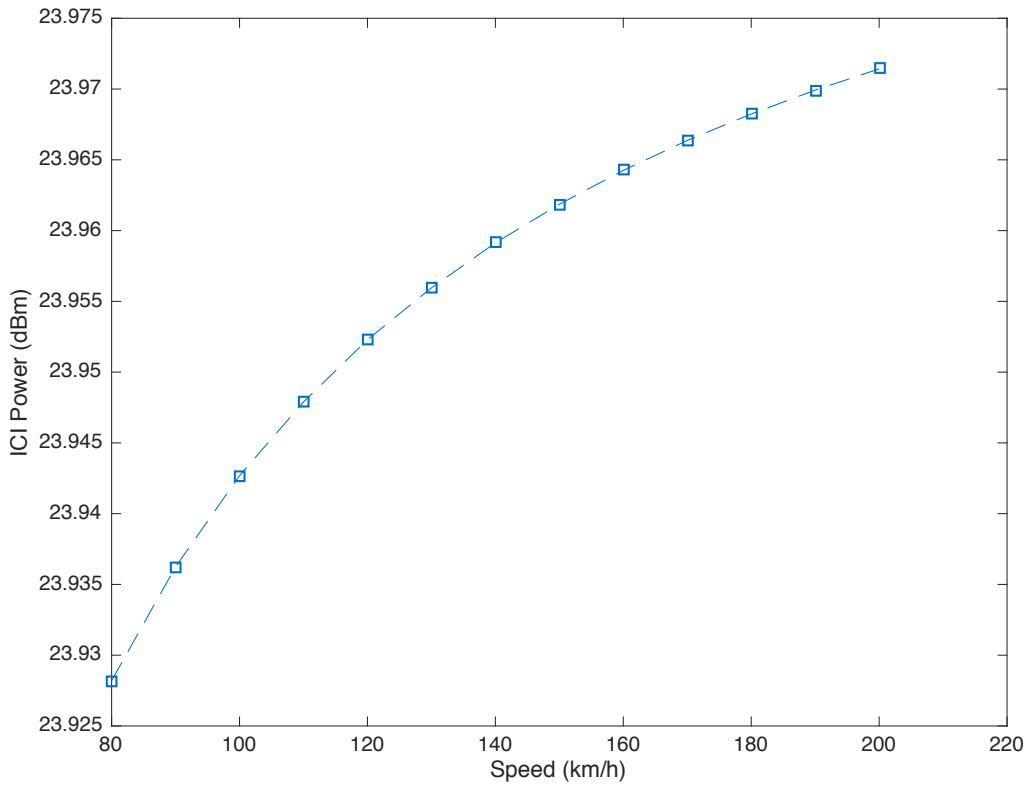

Figure 4. ICI caused by Doppler Shift in AeroMACS

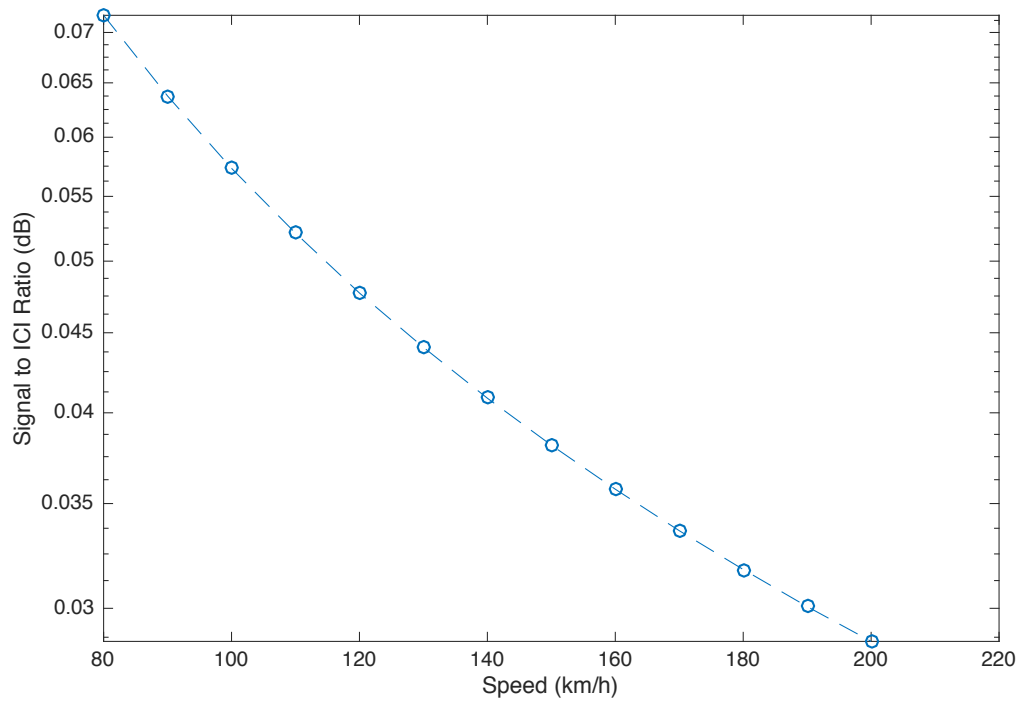

Figure 5. Signal to ICI Ratio

## D. *Coherence Time*

Coherence time, $T_c$, determines fading, that means the time duration over which the channel impulse response is considered to stay constant. Hence, we do not need to update the channel estimation for that period of time. With large $T_c$, the system design will be simpler. No complex channel estimation method or a high load of computation would be necessary.

The equation for coherence time caused by Doppler shift is as follows:

$$T_c = \sqrt{\frac{9}{16\pi {f_D}^2}} = \frac{0.423}{f_D} \qquad (5)$$

Fig. 6 shows the coherence time based on different MS speeds.

To make sure that the communication system has no ICI problem, the subcarrier spacing, Δf, should be more than 5 times the Doppler spread [23]. It is important to note that due to the different paths the signal and its reflections might take, there will be a range of different Doppler shifts at the receiver causing signal fading. This phenomenon is called Doppler spread, which is related to the coherence time.

$$\Delta f > 5 \times \text{Doppler Spread} \qquad (6)$$

Based on the (6), Doppler spread for AeroMACS must be less than 2 kHz. To calculate the coherence time, we can use equation (7):

$$\text{Doppler Spread} = {}^{1}/_{T_c} \qquad (7)$$

Having Doppler spread less than 2 kHz means the coherence time must be greater than 0.5 ms.

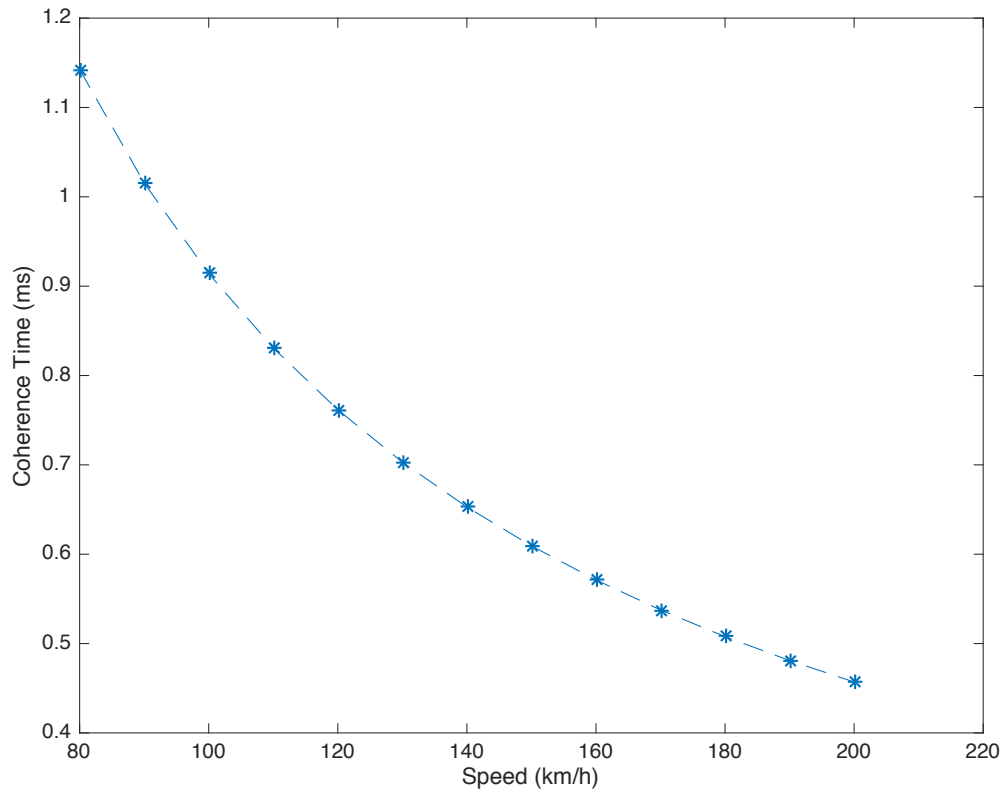

Figure 6. Coherence time versus the speed of aircraft

Finally, using (5), we can calculate the maximum speed of the aircraft supported by AeroMACS that would not cause interference problem. Hence, the aircraft speed should be less than 35.9 m/s or 129.25 km/h.

This gives up a proper upper limit for the UAV's speed, which is sufficient enough for most of the civilian UAV's applications. However, if larger amount of UAV's speed is required, AeroMACS must be able to operate at lower frequencies to avoid interference problems. Another solution to have a higher speed for UAV would be increasing the subcarrier spacing, which helps the system tolerate higher range of Doppler spread.

### *E. Cyclic Prefix Length*

Each symbol transmitted on the physical medium starts with a Cyclic Prefix (CP) part that is repeated at the end of the symbol, as a guard band between the symbols. Thus, if some part of the CP is lost due to symbol spread, it can be recovered from the other end, or to solve the problem of ICI. The CP length is chosen to be a fraction of the symbol length:

$$\mathrm{CP} = G \times T_s \tag{8}$$

Here, $T_s$ is the symbol period, and $G$ is the $\mathrm{CP}/T_s$ ratio. The choice of $G$ is made according to channel parameters. It is usually in the form of $1/2^k$, where $k$ can be any number between 2 to 8.

One of the network parameters that is related to CP is the data rate. The data rate R can be calculated as follows:

$$R = \frac{N_s \times b}{\mathrm{CP} + T_s} \tag{9}$$

Here, b is the number of bits per symbol, $T_s$ is the symbol period, and $N_s$ is the number of subcarriers. Note that longer the CP, the lower is data rate. However, longer CP also minimizes loss due to inter-symbol interference:

$$\mathrm{SNR}_{loss} = -10\ \log_{10}(1 - \frac{\mathrm{CP}}{T_{frame}}) \tag{10}$$

Here $T_{frame}$ is the summation of $T_s$ and the length of the CP. The larger CP results in less propagation loss on data link. As a result, choosing the optimal CP is also a trade-off.

CP is usually chosen to be larger than the maximum delay spread of the channel. Based on studies and experiments reported in [24], in an airport, approximately 10.2 $\mu$s of delay spread happens for a distance of 10,000 feet (which is equal to 3.048 km). Therefore, for 3 km coverage of an AeroMACS cell, the maximum delay spread would be about 10 $\mu$s. Since the CP has to be greater than 10 $\mu$s, we would have, $G \times 102.4\ \mu s > 10 \mu s$, here 102.4 μs is the AeroMACS symbol period. Hence, $G$ must be greater than about 0.1 and in form of $1/2^k$. So, the CP in the AeroMACS has chosen to be about $1/8\, T_s$= 12.8 $\mu$s, with $k = 3$.

## V. SUMMARY

AeroMACS is going to be an important part of the FAA's NEXTGEN program, which will be fully implemented by 2025 [25]. This standard was designed initially for stationary communications at the airports, but soon it got the attention of researchers and investors and now is one of the leading aviation standards. AeroMACS is currently a strong candidate to be used as an aviation standard for the UAV data link.

In this paper, we presented a theoretical analysis of the OFDMA structure of the AeroMACS physical layer. For each feature, the advantages and disadvantages of AeroMACS for different UAV's application were discussed. We also showed the limitations of the aircraft speed, which is due to the effect of Doppler shift. We discussed different parameters, such as subcarrier spacing, maximum coverage, inter-carrier interference, coherence time and cyclic prefix length.

## VI. ACKNOWLEDGEMENT

Dr. Marcio A. Teixeira has been supported under the grant#2017/01055-4 São Paulo Research Foundation (FAPESP).